\documentclass[twocolumn,showpacs,preprintnumbers,amsmath,amssymb]{revtex4}
\usepackage{graphicx}
\usepackage{dcolumn}
\usepackage{bm}
\usepackage{braket}
\usepackage{amsmath}
\def\degC{\kern-.2em\r{}\kern-.3em C}

\begin{document}


\title{Spectroscopy and frequency measurement of the $^{87}$Sr clock transition by laser linewidth transfer using an optical frequency comb}

\author{Daisuke Akamatsu}\email{d-akamatsu@aist.go.jp}
\author{Hajime Inaba}
\author{Kazumoto Hosaka}
\author{Masami Yasuda}
\author{Atsushi Onae}
\author{Tomonari Suzuyama}
\author{Masaki Amemiya}
\author{Feng-Lei Hong}

\affiliation{
National Metrology Institute of Japan (NMIJ), National Institute of Advanced Industrial Science and Technology (AIST), Tsukuba Central 3, Ibaraki 305-8563, Japan}

\date{\today}

\begin{abstract}
We perform spectroscopic observations of the 698-nm clock transition in $^{87}$Sr confined in an optical lattice using a laser linewidth transfer technique. A narrow-linewidth laser interrogating the clock transition is prepared by transferring the linewidth of a master laser (1064 nm) to that of a slave laser (698 nm) with a high-speed controllable fiber-based frequency comb. The Fourier-limited spectrum is observed for an 80-ms interrogating pulse. We determine that the absolute frequency of the 5s$^{2}$ $^{1}$S$_{0}$ - 5s5p $^{3}$P$_{0}$ clock transition in $^{87}$Sr is 429 228 004 229 872.0 (1.6) Hz referenced to the SI second.
\end{abstract}

\maketitle

The precise measurement of time and frequency is of great interest for a wide range of applications, including fundamental science and metrology. Optical frequency standards have been extensively studied and great improvements have been made as regards their frequency stability and uncertainty over the decades. Optical clocks have been realized with various ions and atoms including Al$^{+}$\cite{Al+}, Hg$^{+}$\cite{Hg+}, Sr\cite{Sr}, and Yb\cite{Yb-NMIJ, Yb-NIST}. Since some optical clocks have already outperformed the best microwave clocks, they are being studied by the International Committee for Weights and Measures (CIPM) with a view to redefining the second. As with several single ion optical clocks, both Sr and Yb optical lattice clocks were recommended by the CIPM as secondary representations of the second (candidates for the redefinition of the second) \cite{secondary}. As for the Sr optical lattice clock, the absolute frequency of the clock transition was measured by groups at JILA (Boulder) \cite{Sr-JILA}, SYRTE (Paris) \cite{Sr-SYRTE,Sr-SYRTE2}, UTokyo-NMIJ \cite{Sr-Tokyo-NMIJ}, PTB (Braunschweig) \cite{Sr-PTB}, and NICT (Koganei) \cite{Sr-NICT,Sr-NICT2}. The CIPM is expecting to receive more measurement results from independent groups to confirm the agreement of the clocks, search for any undiscovered uncertainty factors, and update the recommended clock frequency. Furthermore, the realization of different kinds of high-performance optical clocks would allow us to measure the high precision frequency ratio of clocks, which is important for tests in fundamental physics, such as the temporal variation of the fine-structure constant \cite{ratio}.

In the clock experiments, a clock laser with an ultranarrow linewidth is usually prepared in order to interrogate the clock transition by stabilizing its frequency to a high-finesse optical reference cavity with an ultra-low-expansion (ULE) spacer and mirrors with high reflectivity and a low loss coating. However, thermally induced fluctuations in the constituents of the cavity impose a fundamental limit on the linewidth of the clock laser\cite{Numata}. To reduce the thermal noise, a silicon reference cavity is realized that operates at a cryogenic temperature \cite{cryogenic}. In the same context, a mirror with an AlGaAs multilayer coating has also been studied\cite{AlGaAs}. Although these experiments were not carried out at a clock transition frequency, a frequency comb makes it possible to transfer the linewidth and stability to that at the clock transition frequency. Since a frequency comb can distribute the linewidth of a master laser to slave lasers at arbitrary wavelengths covered by the frequency comb, only a single ultranarrow linewidth master laser is necessary even in an experiment with plural clock transitions.

To measure the clock transition of $^{87}$Sr, the frequency stability of a light source at 729 nm was transferred to a light source at 698 nm with a titanium sapphire laser comb \cite{Sr-NICT}. However, because of the limited feedback bandwidth of the comb, the linewidth of the 729-nm laser was not transferred to the 698-nm laser. In this case, in addition to the stability transfer using the comb, a high-finesse cavity was needed to pre-stabilize the laser at 698 nm. A fiber-based frequency comb (fiber comb) is superior to a titanium sapphire comb in terms of its long-term robustness, which is an important characteristic for frequency measurement experiments. Very recently, a group at PTB reported experimental results on transferring a laser frequency stability at the $10^{-16}$ level with a fiber comb and demonstrated the spectroscopy of Sr in an optical lattice \cite{transfer-PTB}. However, because of the limiting servo bandwidth of the comb (500 Hz), they also needed an extra high finesse cavity for pre-stabilization of the clock laser. The feedback bandwidth of the fiber comb can be increased with an intracavity electro-optic modulator (EOM) \cite{intracavityEOM}, and the increased feedback bandwidth is sufficiently broad to reduce the comb linewidth\cite{Nakajima, Iwakuni}. Using a narrow linewidth fiber comb, the spectroscopy of $^{171}$Yb in an optical lattice was successfully demonstrated based on laser linewidth transfer\cite{Yb-Yasuda, Yb-Inaba}. However, in the previous experiment \cite{Yb-Inaba}, the Fourier limited spectrum was only observed for an interrogation time of up to 40 ms (corresponding to the spectrum linewidth of 20 Hz).

In the present work, we observe the Fourier limited spectrum for an 80 ms interrogation pulse. We also perform the absolute frequency measurement of the 5s$^2$ $^1$S$_0$ $-$ 5s5p $^3$P$_0$ clock transition in $^{87}$Sr with the laser linewidth transfer technique. In this experiment, a narrow-linewidth laser operated at 689 nm for intercombination magneto-optical trapping (MOT) is also prepared with the laser linewidth transfer technique\cite{RedMOT-Daisuke}. This has reduced the work needed to build a 689-nm high-finesse cavity and simplified the optical lattice clock system. Furthermore, the optical frequency comb described in this paper can be employed to prepare clock lasers for both Yb and Sr optical lattice clocks from a common master laser. In such cases, since the frequency noise of the clock lasers is correlated, the measurement stability of a Yb/Sr frequency ratio measurement can be improved by a synchronous interrogation of both clock transitions\cite{synchronise}.

\begin{figure}
\begin{center}
\includegraphics[width=\linewidth]{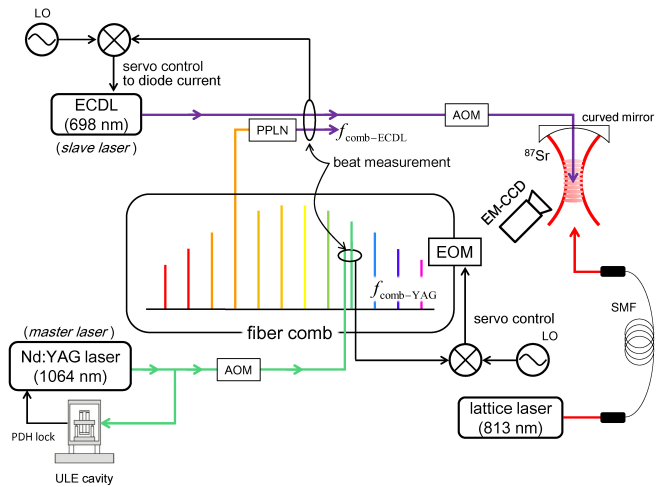}
\end{center}
\caption{Schematic diagram of experimental setup. ECDL, extended cavity diode laser; PPLN, periodically poled lithium niobate; ULE cavity, ultra-low-expansion cavity; PDH lock, Pound-Drever-Hall lock; AOM, acousto-optic modulator; EOM, electro-optic modulator; SMF, single-mode fiber; EM-CCD, electron multiplying charged coupled device. Local oscillators (LO) are microwave sources.}
\label{f1}
\end{figure}

Figure \ref{f1} is a schematic diagram of our experimental setup. The master laser was a monolithic non-planar-ring-oscillator (NPRO) Nd:YAG laser emitting at 1064 nm with a free-running linewidth of several kHz. The frequency of the master laser was stabilized to a high-finesse ULE cavity by using the Pound-Drever-Hall method. The instanteneous linewidth of the master laser was less than 3.5 Hz. However, the linewidth including frequency jitter at an averaging time of several seconds was 10 Hz. We compensated for the linear frequency drift of the master laser (85 mHz/s) caused by the creep of the reference cavity\cite{narrowlaser-Hosaka} by using an acousto-optic modulator (AOM). A fiber comb equipped with an intracavity EOM was used to obtain a broad servo bandwidth for linewidth transfer. The repetition rate of the comb was approximately 43.4 MHz. The carrier-envelope offset frequency was detected using a conventional {\it f}-2{\it f} interferometer and stabilized to an H-maser at our institute. Our fiber-based frequency comb is described in detail elsewhere\cite{Nakajima}. A heterodyne beat between the master laser at around 1064 nm and the comb component ($f_{\mathrm{comb-YAG}}$) was detected to stabilize the repetition rate of the comb using the intracavity EOM. The linewidth of the comb components was reduced to the Hz level owing to the broad servo bandwidth (approximately 1.4 MHz) of the EOM. The clock laser was a homemade extended cavity diode laser (ECDL) with an anti-reflection coated diode laser. The free-running linewidth of the ECDL was approximately 300 kHz. The ECDL generated 20 mW of output light with an injection current of 70 mA. A heterodyne beat note was observed by mixing part of the ECDL output (1 mW) and the second-harmonic comb modes around 698 nm, which were generated by a periodically poled lithium niobate (PPLN) crystal. The ECDL was phase-locked to one appropriate mode of the narrow-linewidth optical comb by employing the direct feedback control to the injection current of the diode laser. The servo bandwidth of the phase locking was approximately 700 kHz. The tight phase lock allowed the linewidth of the comb to be transferred to the ECDL. The remaining output of the ECDL traveled through a double-pass AOM to control the laser frequency and was used for the spectroscopy of the $^1$S$_0$-$^3$P$_0$ transition. The narrow-linewidth comb was also used for preparing the narrow-linewidth cooling laser for the intercombination MOT at 689 nm (not shown in Fig.\ref{f1}) \cite{RedMOT-Daisuke}.

\begin{figure}
\begin{center}
\includegraphics[width=0.8\linewidth]{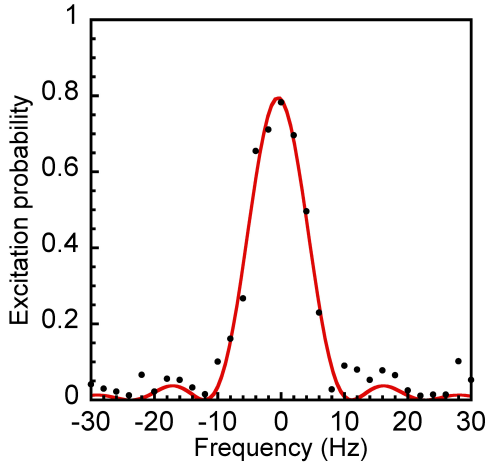}
\end{center}
\caption{Typical spectrum of one Zeeman component ($m_F=-9/2$) of $^{87}$Sr for a $\pi$-pulse of 80 ms, fitted with a sinc function.}
\label{f2}
\end{figure}

The Sr optical lattice clock operates with lattice-confined $^{87}$Sr atoms with nuclear spin $I = 9/2$. After two-stage laser cooling using the $^1$S$_0$-$^1$P$_1$ \cite{BlueMOT-Daisuke} and the $^1$S$_0$-$^3$P$_1$ transitions,\cite{RedMOT-Daisuke} about 10$^3$ atoms were loaded on a vertically oriented one-dimensional (1D) optical lattice. The lattice laser was generated by an ECDL at 813 nm and amplified by a tapered amplifier (TA). A transmission grating was used to remove the amplified spontaneous emission generated by the TA. The polarization axis of the lattice laser was parallel to that of the clock laser and an applied magnetic field during optical pumping and the spectroscopy. The lattice laser frequency was set at 368 554.72(6) GHz, where we confirmed experimentally that the differential AC Stark shift of the clock transition was cancelled. The lattice laser frequency was stabilized to another frequency comb and monitored with a wavemeter. The trap depth of the lattice was $24E_\mathrm{r}$ (recoil energy $E_\mathrm{r}/k_\mathrm{B} = 170\,\mathrm{nK}$). The clock laser was injected from the retro-reflection mirror of the optical lattice and aligned coaxially with the lattice laser so that the clock laser was coupled into the fiber of the lattice laser. The beam diameter of the clock laser was about 310 ${\mathrm{\mu}}$m, which was much larger than that of the lattice beam (180 $\mathrm{\mu}$m). The larger beam waist ensured the homogeneous excitation of the atoms. To reduce line-pulling and collision shifts, the atoms were optically pumped to either of the stretched states $\vert m_F=\pm 9/2\rangle$ by a circularly polarized 25 ms light pulse resonant with the $^1$S$_0$ ($F=9/2$) - $^3$P$_1$ ($F=9/2$) transition. During optical pumping, a homogeneous magnetic field of 91 $\mu$T parallel to the polarization of the lattice and clock lasers was used to define the quantization axis. The magnetic field was kept during the spectroscopy. To observe the clock transition, a first $\pi$-pulse by the clock laser was applied to the stretched state atoms $\vert m_F=\pm 9/2\rangle$ in the optical lattice to excite the atoms from the $^1$S$_0$ state to the $^3$P$_0$ state. To count the number of unexcited atoms, we measured the fluorescence ($I_{\mathrm{unexcited}}$) on the strong $^1$S$_0$-$^1$P$_1$ transition using an electron multiplying charged coupled device (EM-CCD) camera. We then applied a second $\pi$-pulse of 4.5 ms to repump the atoms in the $^3$P$_0$ state to the $^1$S$_0$ state and counted the number of excited atoms by measuring the fluorescence ($I_{\mathrm{excited}}$). Figure \ref{f2} shows a typical Zeeman component ($m_F=-9/2$) signal ( $I_{\mathrm{excited}} /( I_{\mathrm{excited}} + I_{\mathrm{unexcited}})$) obtained by the first $\pi$ pulse of 80 ms. A lineshape of 10 Hz is observed with a high excitation probability. The spectral structure of the $m_F=\pm 7/2$ component due to the imperfect optical pumping was not observed. Therefore we neglected the line-pulling effect in the frequency measurement. For the clock operation, we optically pumped the atoms to each of the stretched states in turn. The interrogation cycle for each stretched state was 2 s. By averaging the frequency of the two states, we could eliminate the first order Zeeman shift and determine the line center of the clock transition. In practice, by probing both sides of the spectrum at its full width at half maximum, the deviation of the clock laser frequency from the atomic resonance was detected and used to stabilize the clock laser frequency.

\begin{figure}[h]
\begin{center}
\includegraphics[width=0.8\linewidth]{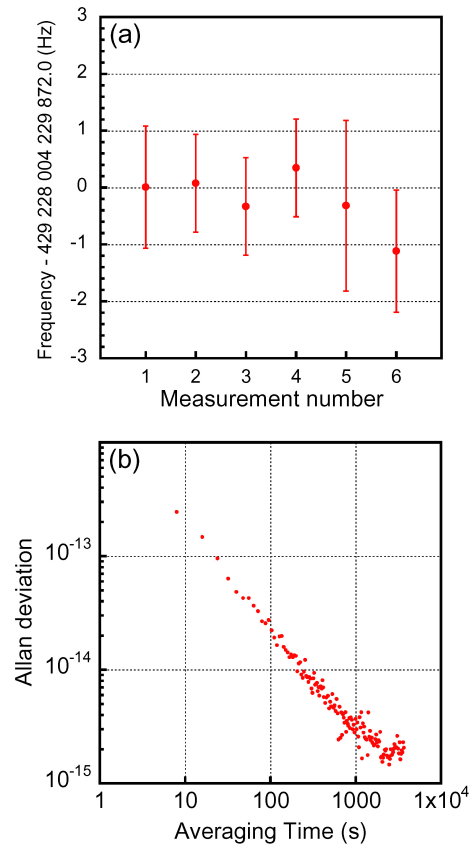}
\end{center}
\caption{(a) Absolute frequency measurements obtained over 5 days. Data shown in this figure do not include systematic corrections. (b) The Allan standard deviation of the frequency measurement against the UTC(NMIJ).}
\label{f3}
\end{figure}

For absolute frequency measurements, the repetition rate of the frequency comb used for the linewidth transfer and the frequency driving the AOM to stabilize the clock laser were recorded by frequency counters, whose microwave references were coordinated universal time UTC(NMIJ). Figure \ref{f3}(a) shows 6 absolute frequency measurements obtained over 5 days (May 23rd to 27th, 2013). Figure \ref{f3}(b) shows a typical Allan standard deviation of the measured absolute frequency (measurement \# 3 in Fig. \ref{f3}(a)) against the UTC(NMIJ). The Allan standard deviation was $2.4\times10^{-13}$ for an 8 s averaging time, and it improved after a 3700 s averaging time to $2.0\times10^{-15}$. We note that the measured Allan deviation follows the stability of the UTC(NMIJ). The uncertainty was given by the Allan standard deviation at the longest averaging time for each measurement (ex. the uncertainty for measurement \# 3 in Fig. \ref{f3}(a) was calculated by using the Allan standard deviation in Fig. \ref{f3}(b)). The standard deviation of the measurements of over 5 days (shown in Fig. \ref{f3} (a)) was calculated to be $9.3\times10^{-16}$, which indicates the measurement statistics in the experiment.

The corrections and uncertainties for the $^{87}$Sr optical lattice clock are listed in Table \ref{t1}. The largest correction was due to the blackbody radiation shift, which was estimated to be $55.3(2.3)\times10^{-16}$ for the temperature of the vacuum chamber (T = 303(3) K)\cite{BBR}.  For the AC Stark shift caused by the lattice laser, the 2nd order Zeeman shift, and the collision shift, we experimentally determined the systematic shift by employing an interleaved scheme, where we varied the parameters to be investigated between two values synchronized with the clock cycle. The AC Stark shift caused by the lattice laser, 2nd order Zeeman shift, and collision shift were determined experimentally to be $0.0(1.8)\times10^{-16}$, $-4.3(1.1)\times10^{-16}$, and $0.0(2.1)\times10^{-16}$, respectively. The AC Stark shift induced by the clock laser was estimated to be $-0.2(2)\times10^{-16}$. The gravitational shift was calculated to be $22.7(6)\times10^{-16}$ using the height of the trapped atoms from the geoid surface of 20.9(5) m. While stabilizing the clock laser to the clock transition, we recorded the excitation ratio and evaluated the servo error to be $0.5\times10^{-16}$. All the above corrections add up to $37.0\times10^{-16}$ with an uncertainty of $3.8\times10^{-16}$(Table \ref{t1}).

The UTC(NMIJ) is compared with TAI via a satellite link at five-day intervals. On the other hand, the Sr lattice clock was operated several hours a day. We conservatively estimate the uncertainty due to the measurement time gap to be $33.0\times10^{-16}$ based on the frequency variation of UTC(NMIJ). The results of the frequency link between the UTC(NMIJ) and TAI can be found in Circular T, which is published on the BIPM web page\cite{CircularT}. From Circular T 305, the correction of the link between UTC(NMIJ) and TAI was $-6.9(13.1)\times10^{-16}$. The correction between TAI and SI of $4.0(2.0)\times10^{-16}$ was also obtained from Circular T 305.

\begin{table}
\caption{Frequency corrections and their uncertainties.}
\label{t1}
\begin{ruledtabular}
\begin{tabular}{lcc}
\hline
Effect & Correction & Uncertainty \\
&(10$^{-16}$)&(10$^{-16}$)\\
\hline
Blackbody radiation & $55.3$ & $2.3$ \\
AC Stark (lattice) & $0.0$ & $1.8$ \\
AC Stark (probe) & $0.2$ & $0.2$ \\
2nd order Zeeman & $4.3$ & $1.1$ \\
Collision & $0.0$ & $2.1$ \\
Gravitation & $-22.7$ & $0.6$ \\
Servo error  & $0.0$ & $0.5$\\ 
\hline
Sr systematics total & $\mathbf{37.0}$ & $\mathbf{3.8}$ \\
\hline
UTC(NMIJ) & -- & $33.0^{*}$ \\
UTC(NMIJ) - TAI & $-6.9$ & $13.1$ \\
TAI - SI & $4.0$ & 2.0 \\
Measurement statistics & &$9.3$\\
\hline
Total & $\mathbf{34.1}$ & $\mathbf{37.0}$ \\
\hline
\end{tabular}
\small
{\footnotesize $^{*)}$The uncertainty is due to the short measurement time. See text.}
\end{ruledtabular}
\end{table} 

\begin{figure}
\begin{center}
\includegraphics[width=\linewidth]{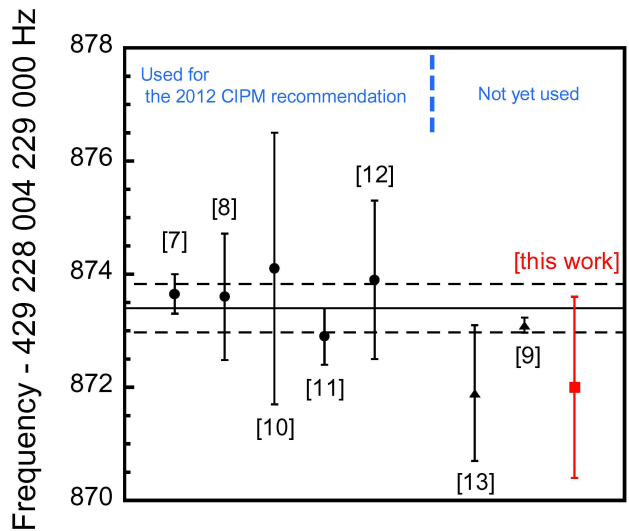}
\end{center}
\caption{Comparison of the absolute frequencies of the $^{87}$Sr clock transition measured by different laboratories. The horizontal solid and broken lines represent the recommended value and its uncertainty for $^{87}$Sr by the CIPM in 2012. The solid circles represent the data used for the 2012 CIPM recommendation. The solid triangles represent the data not yet reported to the CIPM. The red solid rectangle represents the result of this work.}
\label{f4}
\end{figure}

The total systematic uncertainty of the absolute frequency measurement in Table \ref{t1} is $37.0\times10^{-16}$ (1.6 Hz), including the measurement statistics. The inclusion of the corrections in Table \ref{t1} gives a final absolute frequency of 429 228 004 229 872.0 (1.6) Hz referenced to the SI second. Figure \ref{f4} shows the frequencies of the $^{87}$Sr clock transition measured by different laboratories. Our measurement result agrees with measurements performed by different laboratories within the uncertainty. The measured absolute frequency will be reported to the CIPM via the next Consultative Committee for Time and Frequency (CCTF) meeting scheduled for 2015, and will contribute to discussions regarding the updating of the recommended frequency value of the $^{87}$Sr lattice clock, which is one of the secondary representations of the second.

\begin{acknowledgements}
This research receives support from the JSPS through its FIRST Program and JSPS KAKENHI Grant Number 13222778.
\end{acknowledgements}

\end{document}